\documentclass[
    ,final            
  ]
  {aipproc}
\usepackage{bm}

\layoutstyle{8x11single}

\newcommand{\pT}{\mbox{$p_\perp$}}

\newcommand{\vs}{\textit{vs.}}
\newcommand{\sqrts}{\mbox{$\sqrt{s}$}}
\newcommand{\sqrtsNN}{\mbox{$\sqrt{s_{NN}}$}}
\newcommand{\jpsi}{\mbox{$J/\psi$}}
\newcommand{\upsi}{\mbox{$\Upsilon$}}
\newcommand{\upsione}{\mbox{$\Upsilon(\textrm{1S})$}}

\newcommand{\gevc}{\mbox{$\mathrm{GeV}/c$}}

\newcommand{\gevcc}{\mbox{$\mathrm{GeV}/c^2$}}

\newcommand{\pp}{\mbox{$p+p$}}
\newcommand{\PbPb}{\mbox{Pb+Pb}}
\newcommand{\Npart}{\mbox{$N_{\mathrm{part}}$}}

\newcommand{\Zzero}{\mbox{$Z^0$}}
\newcommand{\RAA}{$R_{AA}$}

\begin{document}
\title{Dimuon results in Pb+Pb and \protect\bm{$p+p$} collisions in CMS}

\classification{25.75.-q, 14.40.Pq, 13.20.Gd, 14.70.Hp, 25.75.Nq, 12.38.Mh}
\keywords      {Heavy-ion collisions, Upsilon, $Z^0$ bosons, Quark-Gluon Plasma}

\author{M. Calder\'{o}n de la Barca S\'{a}nchez, on behalf of the CMS Collaboration
\footnote{http://greybook.cern.ch/programmes/experiments/CMS\_details.html
}}{
  address={Physics Department, UC Davis. One Shields Ave, Davis, CA. 95616}
}

\begin{abstract}
The LHC offers unique opportunities for studying the properties of hot QCD matter created in \PbPb\ collisions at extreme temperatures and very low parton momentum fractions. With its high precision, large acceptance for tracking and calorimetry, and a trigger scheme that allows analysis of each minimum bias Pb+Pb event, CMS is fully equipped to measure dimuons in the high multiplicity environment of Pb+Pb collisions. Such probes are especially relevant since they are produced at early times and propagate through the medium, mapping its evolution. The capabilities of the CMS experiment to study dimuon production in pp and Pb+Pb collisions based on the 2010 LHC runs will be reviewed. CMS is able to measure primary and secondary \jpsi, as well as the three \upsi\ states. Quarkonia results at \sqrts=2.76 TeV in \pp\ and \PbPb\ will be presented, including a tantalizing observation of suppression of the \upsi\ excited states. The \Zzero\ boson inclusive and differential measurement as a probe of the initial state will be described.
\end{abstract}

\maketitle


\paragraph{Introduction}
Quantum Chromodynamics (QCD) predicts that at high temperatures and densities strongly interacting matter undergoes a transition from confined hadrons to a color-deconfined state,
known as the Quark-Gluon Plasma.  Relativistic heavy-ion collisions are the only way to access
this state of matter in the laboratory.  One of the key signatures for color deconfinement
is the suppression of heavy quarkonia states.  Originally postulated as the result of color screening of the QCD potential~\cite{Matsui:1986dk}, there has been a whirlwind of activity
indicating that quarkonium is a sensitive probe of additional QGP deconfinement effects, such as Landau damping.  See Ref.~\cite{Brambilla:2010cs} Ch. 5, for a review of Quarkonium in-medium, including descriptions of the suppression observed at SPS and RHIC.  With the first \PbPb\ data from 2010, we discuss the Quarkonia results obtained by CMS on charmonium and bottomonium production via their dimuon decay.

With the energy available in the LHC collisions, it is possible to extend dimuon invariant mass studies in the \PbPb\ system to the previously unreachable $m_{\mu\mu} \approx 100\ \gevcc$ range.  This opens up the study of the \Zzero\ boson in heavy-ion collisions.  This is a probe that is very well understood in \pp\ collisions: the \Zzero\ cross section in \pp\ is in very good agreement with NNLO calculations~\cite{Khachatryan:2010xn}. \Zzero\ production gives heavy-ion experimentalists an electroweak probe which decays into dileptons and will therefore not suffer final-state strong-interaction modifications. It should therefore be very well described by perturbative calculations modulo initial-state effects, such as shadowing of the quark PDFs~\cite{Vogt:2000hp}. We present a summary of our measurements of $\Zzero\rightarrow\mu^+\mu^-$ in the \PbPb\ system.

\begin{figure}[t]
  \includegraphics[height=.3\textheight]{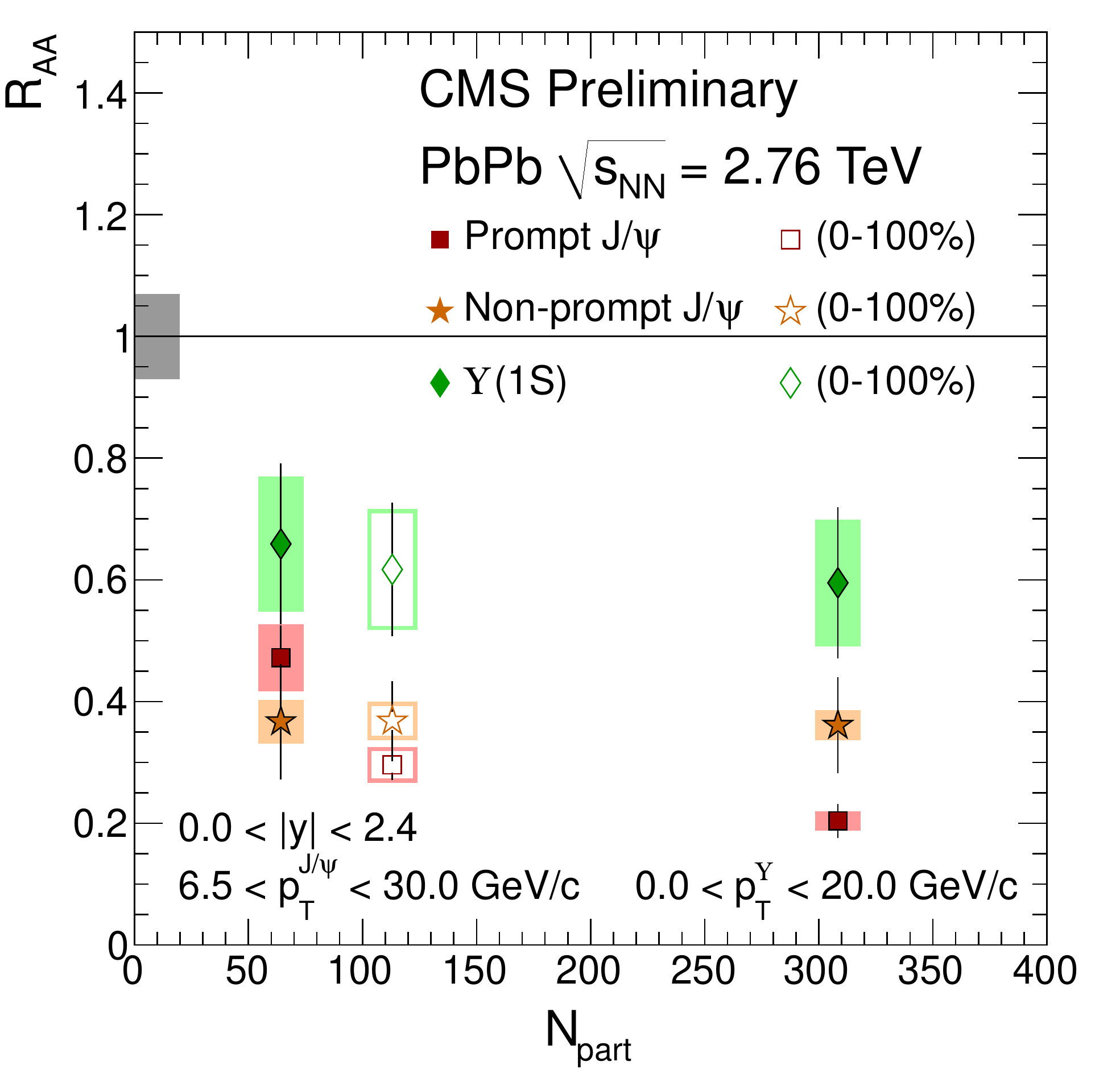}
  \includegraphics[height=.3\textheight]{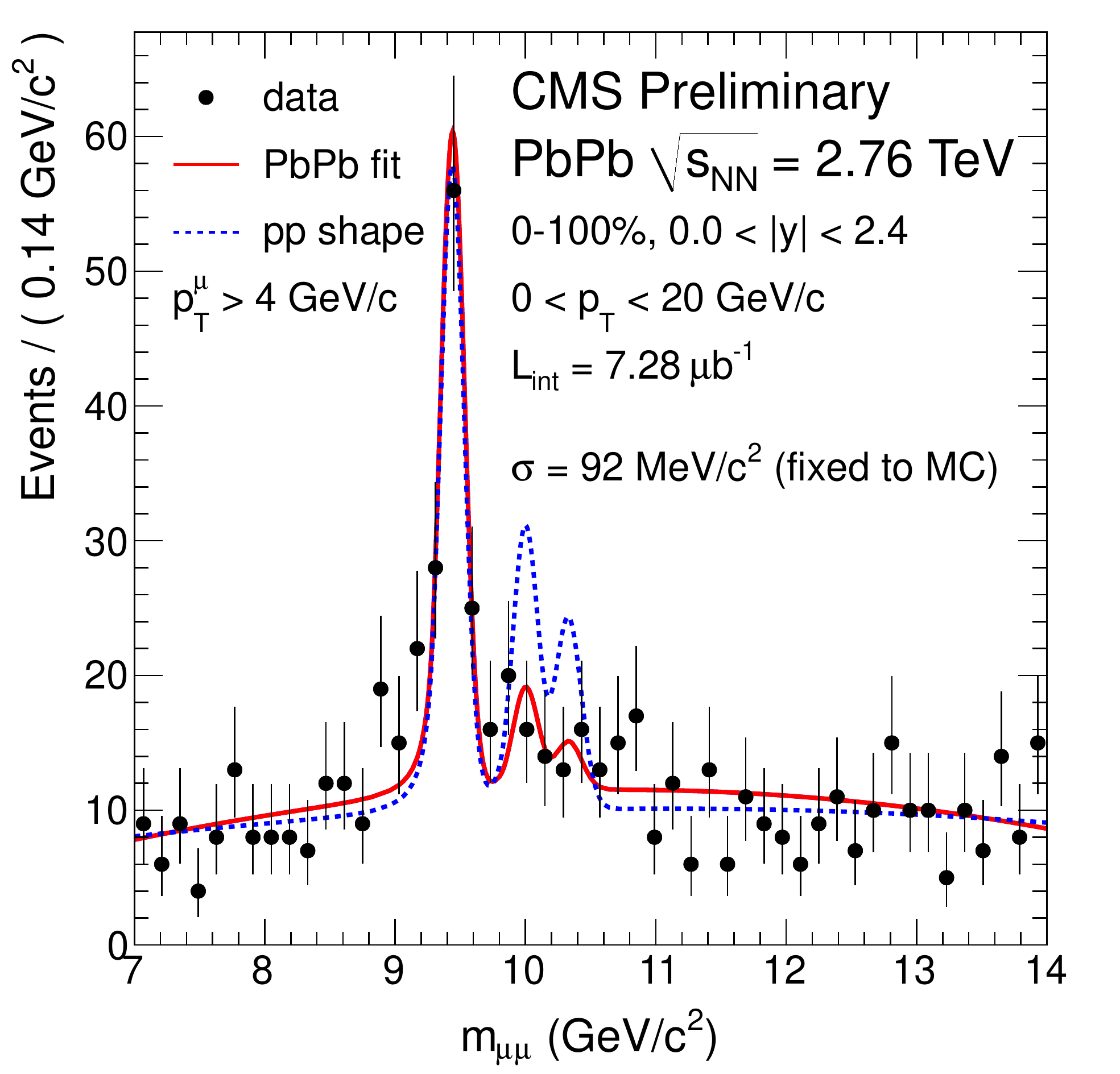}
  \caption{\label{fig:QuarkoniaRAA} (Left) Nuclear modification factor of prompt \jpsi\ (squares),  non-prompt \jpsi\ (stars) and \upsione\ (diamonds) \vs\ centrality (using \Npart). (Right) The dimuon invariant mass distribution in \PbPb\ collisions. The solid line shows the fit to the data. The dashed line is the line shape obtained from \pp\ at the same \sqrts, normalized at the \upsione\ peak.}
\end{figure}

\paragraph{Dimuons in $Pb+Pb$}
The details of the quarkonia analysis can be found in Ref.~\cite{Chatrchyan:2011, QuarkoniaPAS}.  The analysis is based on a dataset of 55 M minimum bias \PbPb\ collisions at \sqrtsNN=2.76 TeV (integrated luminosity of $\mathcal{L}=7.28\ \mu b^{-1}$) taken with the CMS detector~\cite{:2008zzk}. As reference data, a sample of \pp\ collisions at the same \sqrts\ was taken in March 2011 ($\mathcal{L}=225\ nb^{-1}$). In order to reduce the systematic uncertainties when calculating the nuclear modification factor (\RAA), identical algorithms were used to reconstruct the dimuons from quarkonia.  These algorithms closely followed the analysis carried out for the higher energy \pp\ collisions described in Ref.~\cite{Khachatryan:2010yr}, save for more stringent requirements on the pointing of muon tracks to the primary collision vertex.  This significantly reduced the background, but it also reduced the efficiency for reconstructing non-prompt \jpsi s (which come primarily from weak decay of $B$ mesons). Nevertheless, the pointing accuracy of the CMS inner tracker allowed us to measure the decay distance from the vertex, and perform a fit to extract the yield of prompt and non-prompt \jpsi s separately (a first for heavy-ion collisions).  The excellent dimuon mass resolution also allows us to separate the three \upsi\ states.
Figure~\ref{fig:QuarkoniaRAA} shows the \RAA\ for prompt (squares) and non-prompt (stars) \jpsi\ and for the \upsione\ (diamonds).
We observe that all measured quarkonia show suppression with respect to \pp\ collisions. For prompt \jpsi\ in central collisions, we measure \RAA=0.20 $\pm$ 0.03(stat) $\pm$ 0.01(syst). The suppression of the prompt \jpsi\ is expected if one forms a deconfined QGP.  However, suppression is also possible from cold nuclear matter (CNM) effects, including shadowing and comover absorption.  In models which allow \jpsi\ production via statistical recombination of $c-\bar{c}$ pairs, one could enhance \jpsi\ production.   The centrality dependence of the suppression can shed light on the possible contributions of these effects.

The non-prompt \jpsi\ also shows a rather striking suppression (\RAA = 0.36 $\pm$ 0.08(stat) $\pm$ 0.03(syst) in the 20\% most central collisions).  The intriguing aspect of such a result is that, given that we expect non-prompt \jpsi s to be the product of a $B$ meson decay, we might be witnessing energy loss of the $b$ quark in the QGP medium.  We also observe, with limited statistics, a lack of centrality dependence to this suppression. This should provide some food for thought to models of quenching of bottom jets.

The possibility of measuring \upsi\ production to study the QGP~\cite{Gunion:1996qc} has become important in light of the complications surrounding the \jpsi. Bottomonium production is expected to have a much lower recombination contribution due to the lower $b-\bar{b}$ cross section. In addition, \upsi\ suppression by comover absorption is expected to be negligible~\cite{Lin:2000ke}. Therefore, the \upsi\ states are a cleaner probe of deconfinement effects.  One should note that the observed \upsione\ has contributions from higher excited states.  In calculations of bottomonium on the lattice, the higher excited states are seen to melt at lower temperatures than the ground state.  Hence, the observation of suppression of the \upsione\ can be indicative of the dissociation of the excited states.

In order to focus on this effect, we studied the invariant mass distribution in the \upsi\ region for \pp\ and \PbPb\ collisions.  Figure~\ref{fig:QuarkoniaRAA} (right) shows a fit to the \PbPb\ data, together with the line shape extracted from a similar fit to the \pp\ data normalized such that the \upsione\ peaks coincide.  This graphically illustrates that we observe suppression of the excited \upsi\ states in \PbPb\ collisions.  We quantify this  by calculating the double ratio
\begin{equation}
\frac{\upsi(2S+3S)/\upsione|_{PbPb}}{\upsi(2S+3S)/\upsione|_{pp}} = 0.31^{+0.19}_{-0.15}(\textrm{stat}) \pm 0.03 (\textrm{syst}).
\label{eq:UpsDoubleRatio}
\end{equation}
The double ratio method allows the cancelation of important theoretical and experimental uncertainties.  Differences in the choice of initial state PDF, shadowing parameterizations, and renormalization scales will cancel in the single ratios $\upsi(2S+3S)/\upsione$. Experimental acceptance differences, which play a role in the single ratio given that we apply a \pT\ cut on both muons of 4 \gevc, will cancel in the double ratio because this effect will be the same in both systems if we use identical \pT\ cuts.  So the double ratio is insensitive to changes in the overall scale of the cross section and to detector acceptance, making it an exquisite tool to dissect changes in the relative production of excited \upsi\ states between \PbPb\ and \pp.  When studying the statistical significance of the suppression observation, if one assumes that in the absence of a
suppression due to physics mechanisms the double ratio should be unity, then the probability of observing a ratio of 0.31 or below is 0.9\%, i.e., that corresponding
to 2.4$\sigma$ in a one-tailed integral of a Gaussian
distribution. Therefore, our measurement is indicative that the excited states are suppressed in the \PbPb\ system.  The main physics mechanism which is thought to lead to a suppression of the double ratio is the creation color deconfined matter, an expectation based on lattice QCD.  Models which attempt to explain the data using heavy-quark potentials inspired by lattice studies are starting to appear~\cite{Strickland:2011mw}, and we hope these measurements will herald a new era of bottomonium studies of the QGP.

\begin{figure}[t]
\centering
\includegraphics[width=.59\textwidth]{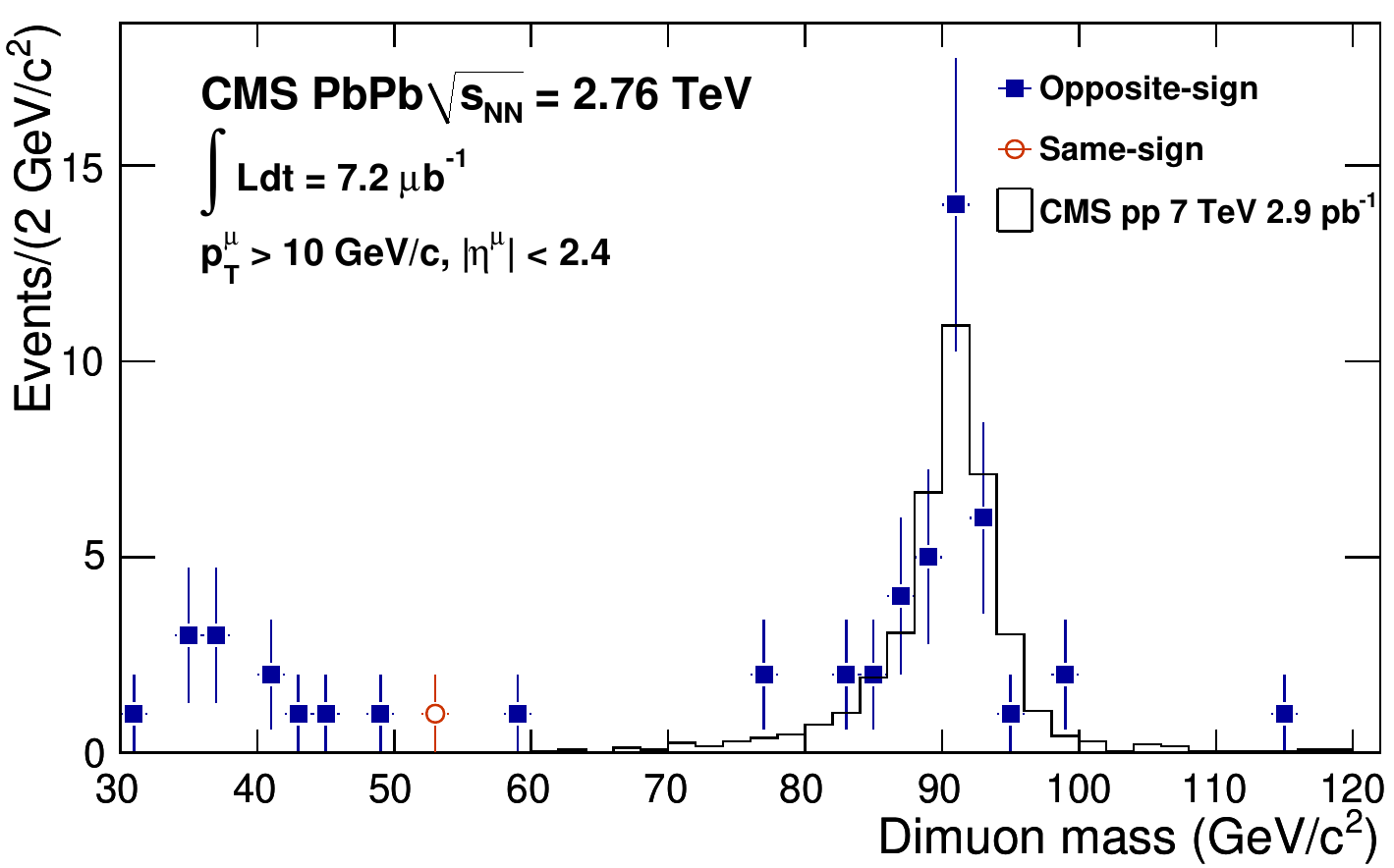}
\quad
\includegraphics[width=.39\textwidth]{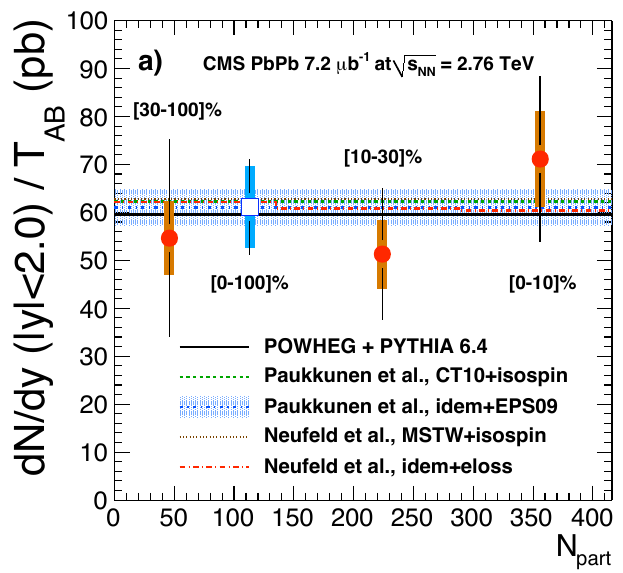}
  \caption{\label{fig:Zboson} (Left) Dimuon invariant mass in \PbPb\ showing a clear Z boson peak. (Right) \Zzero\ yield scaled by $T_{AA}$ \vs\ centrality.}
\end{figure}
The dimuon invariant mass distribution around the \Zzero\ mass from \PbPb\ collisions is shown in Fig.~\ref{fig:Zboson} (left). We see a clear signal with 39 \Zzero\ candidates, and very little background (only one like-sign count).  These were from the same dimuon data sample used for Quarkonia (see Ref.~\cite{Chatrchyan:2011ua} for additional analysis details).  We overlay the line shape of the \Zzero\ measured in \pp\ collisions, illustrating the similarity of the distribution measured in \PbPb.  This is consistent with the expectation that the \Zzero\ should not be strongly modified (in both senses of the word) in \PbPb\ collisions.

Figure~\ref{fig:Zboson} (right) shows a quantitative comparison of the \Zzero\ yield \vs\ \Npart\ with NLO calculations scaled by the nuclear overlap integral, $T_{AA}$, to take into account the nuclear geometry.  Within our uncertainties, we observe no modification of the \Zzero\ cross section, confirming the expectation that it can act as an \textit{in situ}  standard candle of the initial state, in particular of the quark PDFs in the Pb nucleus.  This will be important in future studies of \Zzero\ + jet events in heavy ion collisions, where the \Zzero\ should have identical \pT\ as the recoil jet (to leading order) and thus promise to be a powerful tool to study energy loss in the QGP.

\paragraph{Conclusions}
We presented dimuon results on the production of prompt and non-prompt \jpsi\ and \upsi\ production, all of which display striking suppression patterns.  In particular, the suppression of the \upsi\ excited states promises to be a new window into the physics of color deconfined matter at high temperature. The \Zzero\ boson production is consistent with that in \pp\ after scaling by the nuclear geometry, and thus can be a good reference for the quark distribution function of the Pb nucleus. The author acknowledges the support of the US NSF (Grant No. 0645773).




\bibliographystyle{aipproc}   



\end{document}